\documentclass[aps,showpacs,eqsecnum,nofootinbib]{revtex4}
\usepackage{graphicx}
\usepackage{times,mathptm}
\newcommand{\beq}{\begin{equation}}
\newcommand{\eeq}{\end{equation}}
\newcommand{\bea}{\begin{eqnarray}}
\newcommand{\eea}{\end{eqnarray}}

\renewcommand{\d}{\delta}

\renewcommand{\b}{\beta}

\newcommand{\m}{\mu}
\renewcommand{\r}{\rho}
\newcommand{\bx}{\mbox{\bf x}}
\newcommand{\by}{\mbox{\bf y}}
\newcommand{\bk}{\mbox{\bf k}}
\newcommand{\s}{\sigma}

\newcommand{\E}{{\cal E}}
\newcommand{\V}{{\cal V}}

\newcommand{\tr}{\mbox{Tr}}

\newcommand{\e}{\epsilon}

\newcommand{\oh}{\frac{1}{2}}

\newcommand{\dg}{\dagger}
\newcommand{\non}{\nonumber}

\newcommand{\rf}[1]{(\ref{#1})}
\newcommand{\ra}{\rightarrow}
\newcommand{\pa}{\partial}

\newcommand{\rra}{\right\rangle}
\newcommand{\lla}{\left\langle}
\begin{document}
%
%
\title{Coulomb Energy, Vortices, and Confinement}
\author{Jeff Greensite}
\affiliation{Physics and Astronomy Dept., San Francisco State University,
San Francisco, CA 94117, USA}
\email{greensit@quark.sfsu.edu}
\author{{\v S}tefan Olejn\'{\i}k} 
\affiliation{Institute of Physics, Slovak Academy
of Sciences, SK--845 11 Bratislava, Slovakia}
\email{fyziolej@savba.sk}
\date{March 27, 2003}
\begin{abstract}
We estimate the Coulomb energy of static quarks
from a Monte Carlo calculation of the correlator of timelike link
variables in Coulomb gauge.  We find, in agreement with Cucchieri and
Zwanziger, that this energy grows linearly with distance at large
quark separations.  The
corresponding string tension, however, is several times greater than the
accepted asymptotic string tension, indicating that a state containing
only static sources, with no constituent gluons, is not the lowest energy
flux tube state.  The Coulomb energy is also measured on thermalized lattices
with center vortices removed by the de Forcrand--D'Elia procedure.  We find
that when vortices are removed, the Coulomb string tension vanishes.
\end{abstract}
\maketitle
%
%
\section{Introduction}

   There is an old idea about confinement in Coulomb gauge which was
originally put forward by Gribov \cite{Gribov}, and which has been advocated
in recent years by Zwanziger \cite{Dan}.  The idea is roughly as
follows:  For an SU($N$) gauge theory fixed to ``minimal'' Coulomb gauge,
the path integral is restricted to
the region of configuration space in which the Faddeev--Popov operator
\begin{equation}
         M^{ac} = - \pa_i D_i^{ac}(A)
                = -\nabla^2 \d^{ac} - \e^{abc} A_i^b \pa_i
\end{equation}
has only positive eigenvalues.  The boundary of this region in configuration
space, beyond which $M$ acquires zero or negative eigenvalues, is known as
the ``Gribov Horizon.''  In Coulomb gauge, the inverse of the
Faddeev--Popov operator enters into the non-local part of the Hamiltonian
\begin{equation}
H = \oh \int d^3x ~ (\vec{E}^{a,tr} \cdot \vec{E}^{a,tr}
 + \vec{B}^a \cdot \vec{B}^a)
 + \oh \int d^3x d^3y ~ \r^a(x) K^{ab}(x,y) \r^b(y),
\end{equation}
where $\r^a$ is the (matter plus gauge field) color charge density,
$\vec{E}^{a,tr}$ is the transverse color electric field
operator, and
\begin{equation}
       K^{ab}(x,y) = \left[ M^{-1}(-\nabla^2)M^{-1}
                      \right]_{x,y}^{ab}
\label{Dansway}
\end{equation}
is the instantaneous Coulomb propagator.  The expectation value of the
non-local $\r K \r$ term in the Hamiltonian gives us the Coulomb energy.
Now, since the dimension of configuration space is very large, it is
reasonable that the bulk of configurations are located close to the horizon
(just as the volume measure $r^{d-1} dr$
of a ball in $d$-dimensions is sharply peaked near the radius of the
ball).  Since it is the inverse of the $M$ operator which appears in
the Coulomb energy, it is possible that the near-zero eigenvalues of
this operator will enhance the magnitude of the energy at large quark
separations, possibly resulting in a confining potential at large
distances.  It may be, then, that the static quark potential is simply
the Coulomb potential.  In a diagrammatic  analysis, the area-law falloff
of large timelike Wilson loops in Coulomb gauge
would be obtained by exponentiating
instantaneous one-gluon exchange, going like
\begin{equation}
          D_{00}(\bk,t) \sim {A\over |\bk|^4} \d(t)
\end{equation}
at small $|\bk|$.

   An objection that can be raised to this scenario is that if the
confining potential is coming just from one-gluon exchange, it would be
very hard to understand the origin of string-like behavior indicated by
roughening \cite{rough} and the L\"uscher term \cite{lush}, for which
there is now solid numerical evidence \cite{lush1}.
On the other hand, Zwanziger \cite{Dan1}
has pointed out that the Coulomb energy
of static sources is an upper bound on the static potential.  This
means that if the static potential is confining, then so is the
Coulomb potential.  It then seems economical to identify the two,
as suggested by recent data on the Coulomb propagator reported by
Cucchieri and Zwanziger \cite{ZC}.

   In this article we calculate, by Monte Carlo techniques, the
equal-times correlator of timelike link variables in Coulomb gauge.
The logarithm of this quantity, in the continuum limit, is the Coulomb
energy.  We find, in agreement with Ref.\ \cite{ZC}, that the Coulomb
energy rises linearly with distance at large quark separations.  On
the other hand, we also find that the slope is far too high to
identify the Coulomb string tension with the usual asymptotic string
tension.  We will discuss the relevance of this result to the gluon
chain proposal of Thorn and one of us (J.G.) \cite{gchain}, and in
connection with some related comments of 't~Hooft~\cite{tH}.

   A second motivation of this article is to find out if there is a
connection between the confining Coulomb potential and the center
vortex confinement mechanism.  We have therefore repeated our
calculation of the Coulomb energy, in thermalized lattice
configurations, with
vortices removed by a procedure introduced by de Forcrand and D'Elia
\cite{dFE}.  We find that vortex removal also removes the Coulomb
string tension.

%
%
\section{Link Correlators and the Coulomb Energy}

   We begin by defining the correlator, in Coulomb gauge, of two
timelike Wilson lines
\begin{equation}
      G(R,T) =  \langle \tr[ L^\dg(\bx,T) L(\by,T)] \rangle,
\end{equation}
where $R=|\bx - \by |$ and
\begin{equation}
       L(\bx,T) = P\exp\left[ i\int_0^T dt A_0(\bx,t) \right].
\end{equation}
Wick rotation to Euclidean time is understood.
This correlator represents the creation, at time $t=0$, of two static
sources in a (global) color singlet state separated by a spatial distance $R$.
This is a physical state in Coulomb gauge, denoted by $\Psi_{qq}$,
and can be represented by
massive quark/antiquark creation operators $\overline{q}(\bx) q(\by)$
acting on the true vacuum state $\Psi_0$. The color sources propagate
for a time $T$, and are then annihilated.
From the existence of a transfer matrix we have
\begin{equation}
        G(R,T) = \sum_n | c_n |^2
                      e^{-E_n T},
\label{G}
\end{equation}
where the sum is over all states $|n\rangle$ that have non-vanishing
overlap $c_n$ with $\Psi_{qq}$, and $E_n$ is the energy of the $n$-th state
above the vacuum energy.

   The energy expectation value of the state $\Psi_{qq}$, above the
vacuum energy, consists of an $R$-independent self-energy term
$E_{se}$, plus an $R$-dependent potential.  In this state, an
$R$-dependent energy can only arise from the expectation value of the
non-local Coulomb term in
the Hamiltonian, so the $R$-dependent part of the energy is the Coulomb
potential between static sources $V_{coul}(R)$.
We have
\begin{equation}
       \E = \lla \Psi_{qq}|H|\Psi_{qq}\rra -
              \lla \Psi_0|H|\Psi_0\rra
          = V_{coul}(R) + E_{se}.
\end{equation}
Defining
\begin{equation}
        \V(R,T) = -{d \over dT} \log[G(R,T)]
\end{equation}
it is easy to see that \cite{GH}
\begin{equation}
        \E = \lim_{T\ra 0} \V(R,T),
\label{gh}
\end{equation}
while
\begin{equation}
        E_{min} = \lim_{T\ra \infty} \V(R,T)
                = V(R) + E'_{se},
\end{equation}
where $E_{min}$ is the minimum energy, above the vacuum energy,
of the quark-antiquark system,
$V(R)$ is the usual static quark potential, and $E'_{se}$ is an
$R$-independent self-energy.  The minimum energy state, which dominates
the sum over states in Eq.\ \rf{G}, is expected to represent a flux tube
in its ground state, stretching between the static sources.

   The self-energies $E_{se}$ and $E'_{se}$ are regulated when the gauge 
theory is formulated on the lattice.  If the static quark 
potential is confining, then these self-energies must be negligible, 
at sufficiently large $R$, compared to the static potential.  
Then, from the fact that $E_{min}\le \E$, if follows that
\begin{equation}
         V(R) \le V_{coul}(R),
\label{vv}
\end{equation}
as first pointed out by Zwanziger \cite{Dan1}. This means that if
the static quark potential is confining, the instantaneous Coulomb potential
is also confining.

   With a Euclidean lattice regularization, we define
\begin{equation}
        L(\bx,T) = U_0(\bx,a) U_0(\bx,2a)...U_0(\bx,T)
\end{equation}
as the appropriate ordered product of link variables, and
\begin{equation}
        V(R,T) = {1\over a} \log\left[{G(R,T)\over G(R,T+a)}\right],
\end{equation}
where $a$ is the lattice spacing.  The identity \rf{gh} then holds
only in the continuum limit, i.e.
\begin{equation}
        \lim_{\b \ra \infty} V(R,0) = V_{coul}(R) + \mbox{const.},
\label{Vcoul}
\end{equation}
while we still have, at any $\b$,
\begin{equation}
        \lim_{T \ra \infty} V(R,T) = V(R) + \mbox{const.},
\label{Vstat}
\end{equation}
where the constants are self-energies, and
\begin{equation}
       V(R,0) = -{1\over a}\log[G(R,a)].
\end{equation}
By calculating $V(R,0)$ via lattice Monte Carlo, we may address
several questions:
\begin{enumerate}
\item Does $V(R,0)$ (and hence the Coulomb potential) increase
linearly with $R$ at large $\b$?
\item If so, does the associated Coulomb string tension $\s_{coul}$ match the
usual asymptotic string tension $\s$ of the static quark potential?
\item If center vortices are removed from thermalized lattice configurations,
what happens to the measured Coulomb potential?
\end{enumerate}

%
%
\section{Numerical Results}

  From here on we will work in lattice units, with $a=1$. Lattice sizes
used in numerical simulations are $16^4, 16^4, 20^4$, and $24^4$
at $\beta=2.2, 2.3, 2.4$, and $2.5$ respectively. Data points are derived
from 500 configurations separated by 50 sweeps at each $\b$.

\begin{figure}[tbp]
\centerline{\scalebox{1.0}{\includegraphics{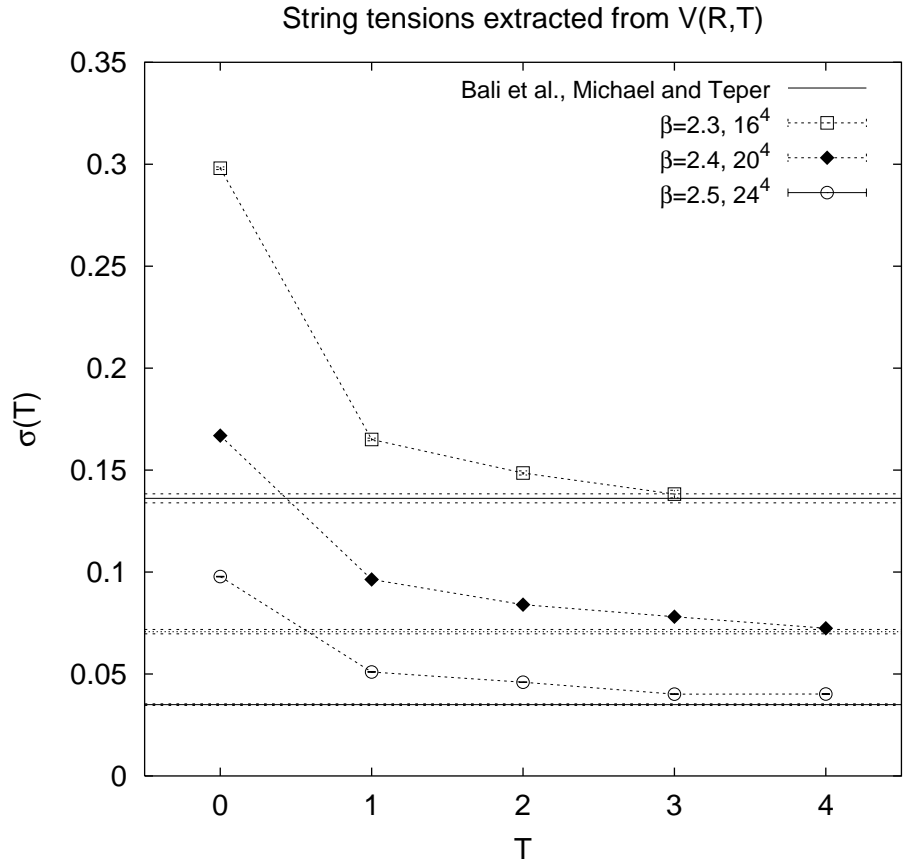}}}
\caption{Falloff of $\s(T)$ with increasing $T$ at $\b=2.3,2.4,2.5$.
Solid lines
indicate the accepted values of the asymptotic string tension at each
$\b$ value, with dashed lines indicating the error bars.}
\label{sigmavsT}
\end{figure}

\begin{figure}[tbp]
\centerline{\scalebox{1.0}{\includegraphics{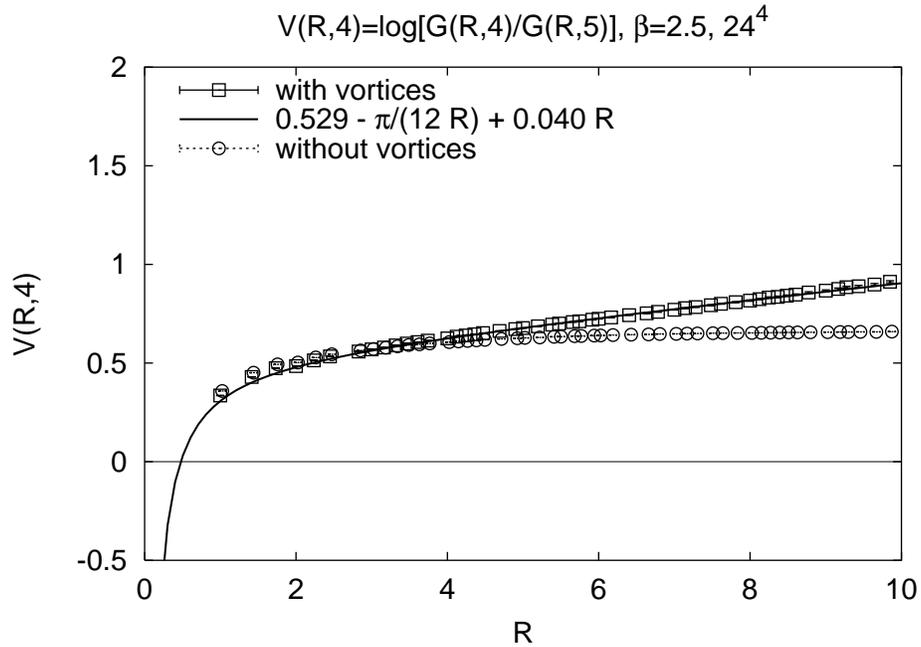}}}
\caption{$V(R,4)$ at $\b=2.5$.  The ``without vortices'' data points
are obtained on lattices with vortices removed by the de Forcrand--D'Elia
procedure \cite{dFE}.} 
\label{v4_2p5}
\end{figure}

\begin{figure}[tbp]
\centerline{\scalebox{1.0}{\includegraphics{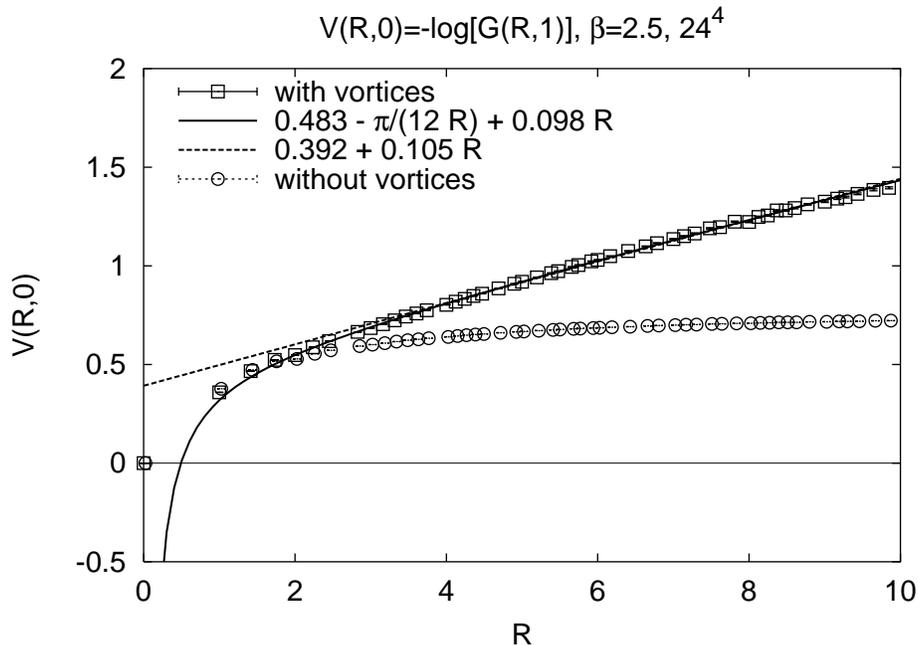}}}
\caption{$V(R,0)$ at $\b=2.5$, which is related [Eq.\ \rf{Vcoul}]
to the Coulomb energy. The solid (dashed) line is a fit to a linear
potential with (without) the L\"uscher term.} 
\label{v0_2p5}
\end{figure}

  First, as a check of Eq.\ \rf{Vstat}, it is useful to verify that
the string tension $\s(T)$ extracted from $V(R,T)$
approaches the accepted asymptotic string tension at large $T$,
where the link correlator $G(R,T)$ is computed in lattice
Coulomb gauge (implemented by the over-relaxation technique).
String tensions $\s(T)$ are extracted from a fit of $V(R,T)$ to
the form
\begin{equation}
        V(R,T) = c(T) - {\pi \over 12 R} + \s(T) R
\label{fit}
\end{equation}
in the range $R_{min}\le R \le R_{max}$, where we have used
$R_{min}=3,3,4,4$, and $R_{max}=5,6,8,10$ at $\b=2.2,2.3,2.4,2.5$
respectively.  The results for $\s(T)$ vs.\ $T$ at $\b=2.3-2.5$
are shown in Fig.~\ref{sigmavsT}.  The data does appear to converge
towards the accepted asymptotic string tension as $T$ increases.  The
data for $V(R,4)$ at $\b=2.5$ is shown in Fig.~\ref{v4_2p5} (data points
marked ``without vortices'' will be explained below).\footnote{In order 
to avoid multiply overlapping, overcrowded 
symbols, not all available data points are displayed in Figs.~\ref{v4_2p5}
and \ref{v0_2p5}.}  In this case
$\s(4)=0.0402(2)$, which can be compared to the accepted asymptotic string
tension $\s=0.0350(4)$ at this coupling \cite{Bali}.  We note that
timelike Wilson line correlators, in a physical gauge,
have been used previously to compute static potentials, most recently
by de Forcrand and Philipsen \cite{dFP} in connection with adjoint
string-breaking.  The static ground-state potential is obtained 
from the asymptotic correlators at large $T$, but in connection with
the Coulomb potential we are interested in the opposite, small $T$ limit.

   As already explained, the Coulomb potential is obtained at large
$\b$ from $V(R,0)$.  At all four values of $\b$ that we have used in
our simulations, $V(R,0)$ is clearly a linear function of $R$ at large
$R$, and we see no reason that this behavior would change as $\b$ is
increased.  This means that the instantaneous Coulomb potential is
also linear at large $R$, and the first question
posed at the end of Section 2 can be answered affirmatively,
in agreement with Cucchieri and Zwanziger \cite{ZC}.
Our data for $V(R,0)$ at $\b=2.5$ is shown in 
Fig.~\ref{v0_2p5}.

\begin{figure}[tbp]
\centerline{\scalebox{1.0}{\includegraphics{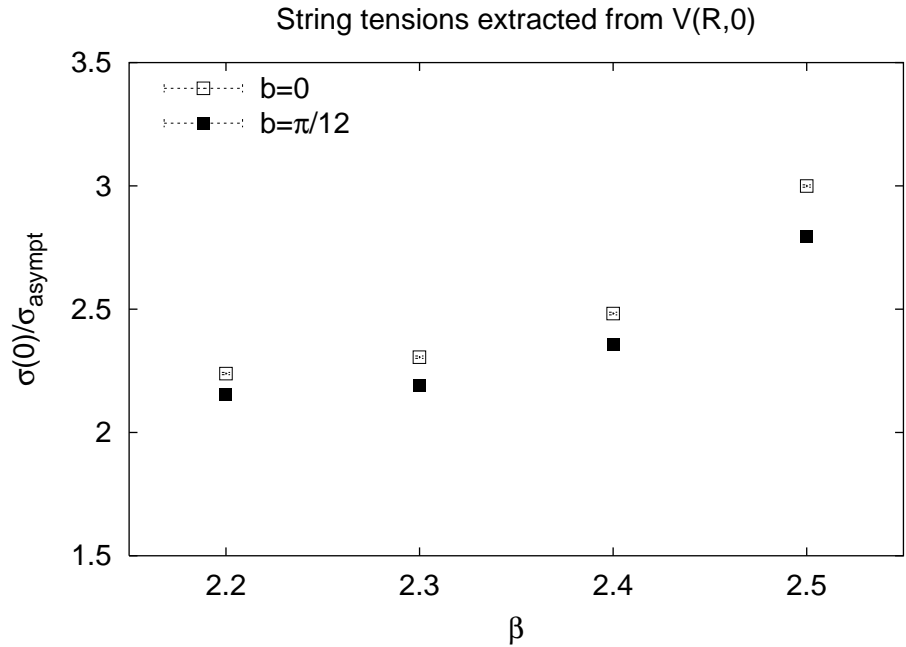}}}
\caption{The ratio $\s(0)/\s$, at various $\b$, from fits which include
($b=\pi/12$) or do not include ($b=0$) the L\"uscher term.  This ratio should
equal the ratio of the Coulomb to asymptotic string tension in
the $\b\ra \infty$ limit.}
\label{sigmavsbeta}
\end{figure}

   On the other hand, there is no indication that
$\s(0)\approx \s(\infty)$ at large $\b$, which is what is required if the
Coulomb string tension $\s_{coul}$ would agree with the usual asymptotic string
tension $\s$.  If anything, there is the opposite tendency as $\b$ increases.
In Fig.~\ref{sigmavsbeta} we plot the ratio $\s(0)/\s$ as a function
of $\b$.  We have obtained $\s(0)$ from two different fits, with
and without the L\"{u}scher term $-\pi/12R$.  It makes sense to include the
L\"{u}scher term at large $T$, but it is a little hard to see how such a
term, derived from string-like fluctuations, would originate due to
instantaneous one-gluon exchange, which is the origin of the Coulomb force.
So we have also extracted $\s(0)$ from a fit in which the L\"{u}scher term
is dropped in Eq.\ \rf{fit}.  In any case the ratios $\s(0)/\s$,
extracted from fits with and without the L\"uscher term, are not
much different, and both results are shown in Fig.~\ref{sigmavsbeta}.
These ratios show a tendency to increase with $\b$, and we do not really
see convergence to a finite value, at least in this range of $\b$, and
certainly no evidence that the ratio converges to unity.  If the rise in
$\s(0)/\s$ is monotonic in $\b$, then assuming the ratio converges at all,
it is unlikely to be less than $\s(0)/\s = 3$.  The data therefore
appears to give a negative answer to the second question posed at the
close of the previous section.  Our results are not compatible with
$\s_{coul} \approx \s$, and we differ in this respect from the conclusions
of Ref.\ \cite{ZC}.

%
%
\subsection{Vortex Removal}

    The idea that confinement is entirely due to the Coulomb potential,
arising from the instantaneous part of the $A_0 A_0$ propagator,
has certain problematic features.  Apart from the issue of the
L\"{u}scher term, it is not entirely clear
how the Coulomb propagator would explain the string
tension of spacelike Wilson loops, which are constructed from the
transverse gluon field.\footnote{We thank D.\ Zwanziger for this comment.}
An alternative explanation of the confining
force in terms of center vortices has been extensively investigated
(cf.\ the review in Ref.\ \cite{rev} and references therein), and this
mechanism can be invoked to obtain an area law falloff for spacelike loops
as well as timelike loops.  It is
interesting, then, to ask if there is some relation between the linear
confining Coulomb potential found in Coulomb gauge, and the effects of
center vortices identified by center projection in an adjoint gauge.

   To study this question, we adopt the ``vortex removal'' method
devised by de Forcrand and D'Elia \cite{dFE}.  In SU(2) lattice
gauge theory, a thermalized lattice is fixed to maximal center gauge,
and then the links are modified according to the rule
\begin{equation}
       U_\m(x) \ra U'_\m(x) = \mbox{signTr}[U_\m(x)]\;U_\m(x).
\end{equation} The modified configuration is still in maximal center gauge,
but has no vortices upon center projection.  The procedure can be
visualized as placing thin center vortices in the middle of thick
center vortices; the effect of the two types of vortices on large
loops cancel out. It is well known that the string tension of
Wilson loops in the modified configuration vanishes \cite{dFE}.
For present purposes we gauge fix the modified $U'_\m(x)$
configuration to Coulomb gauge, and compute $V(R,T)$ as before.
The result, labelled ``without vortices'' in Figs.~\ref{v4_2p5}
and \ref{v0_2p5}, is that the string tension $\s(T)$ vanishes at
every $T$ and every $\b$.  There is no Coulomb string tension, and
no asymptotic string tension, when center vortices are
removed.\footnote{The effect of vortex removal on Landau gauge
propagators, and its possible implications for confinement, has
been investigated by K.\ Langfeld \textit{et al.} in Ref.~\cite{Langfeld}.}

%
%
\section{Coulomb Energy and the Gluon Chain Model}

   We have alluded several times to the string-like behavior of the
QCD flux tube, which manifests itself both in the phenomenon of
``roughening,'' i.e.\ the logarithmic growth of flux tube thickness
with quark separation, and also by the presence of the L\"uscher
term in the static potential.  It is not obvious how
this string-like behavior would be obtained from instantaneous
one-gluon exchange, even given that such an exchange generates a linear
confining potential.  Our data also indicates that the purely
Coulombic force, at long range, may be several times greater than the actual
asymptotic force between static quarks.  We would
like to suggest that the two issues are related: the Coulombic force
is lowered to the true asymptotic force by constituent gluons in the
QCD flux tube, and the fluctuations of these gluons in transverse
directions accounts for the string-like phenomena.  The general picture
is known as the ``gluon chain model,'' advocated by
Thorn and one of us (J.G.) in Ref.\ \cite{gchain}.

   The gluon chain model is motivated by the fact that a time-slice of
a high order planar diagram for a Wilson loop reveals a sequence of gluons,
with each
gluon interacting only with its nearest neighbors in the diagram.  This
suggests that the QCD string might be regarded, in a physical gauge, as
a ``chain'' of constituent gluons, with each gluon held in place by
attraction to its two nearest neighbors in the chain.  The linear potential
in this model comes about in the following way:  As heavy quarks separate,
we expect that the Coulombic interaction energy increases rapidly due to
the running coupling.  Eventually it becomes
energetically favorable to insert a gluon between the quarks, to reduce
the effective color charge separation. As the quarks continue to separate,
the process repeats, and we end up with a chain of gluons.  The average
gluon separation $d$ along the axis joining the quarks is fixed, regardless
of the quark separation $R$, and the total energy of the chain is the energy
per gluon times the number $N=R/d$ of gluons in the chain, i.e.
\begin{equation}
       E_{chain} \approx N E_{gluon} = {E_{gluon}\over d} R = \s R,
\end{equation}
where $E_{gluon}$ is the (kinetic+interaction) energy per gluon, and
$\s=E_{gluon}/d$ is the asymptotic string tension.  In this picture, the
linear growth in the number of gluons in the chain is the origin of the
linear potential.  A typical gluon-chain state would have the form
\bea
  \Psi_{chain}[A] &=& \overline{q}^{a_1}(\bx)
   \Bigl\{\int d\bx_1 d\bx_2 ... d\bx_N
   ~\psi_{\m_1 \m_2 ... \m_N}(\bx_1,\bx_2,...,\bx_N)
\non \\
    & &  A^{a_1 a_2}_{\m_1}(\bx_1) A^{a_2 a_3}_{\m_2}(\bx_2) ...
        A^{a_N a_{N+1}}_{\m_N}(\bx_N) \Bigr\} q^{a_{N+1}}(\by)
        \Psi_0[A]
\label{gluon_chain}
\eea
where $\psi$ is a ``string-bit'' wavefunction correlating positions of
the $N$ constituent gluons in the chain, and is to be determined by
minimizing the energy of the chain.  This minimization only involves
the interaction of neighboring gluons separated by an average distance $d$,
rather than the direct interaction of quarks separated by
a very large distance $R$.  It was shown in Ref.\ \cite{gchain}, on the
basis of a simplified quantum-mechanical model, that gluon-chain states
can plausibly account for the observed string-like behavior of
the QCD flux tube.

   However, as pointed out last year by 't Hooft \cite{tH},
there are certainly states in the Fock space, containing the static
sources, which are not gluon chain
states.  If the gluon chain scenario has any validity, then it must be
that the interaction energy of these non-chain states is also confining,
but of much higher energy than the gluon chain states.

   This is where the numerical result of the last section, which indicates
that $\s_{coul}>\s$, becomes relevant.
The simplest ``no-chain'' state
is a state with no constituent gluons at all, i.e.
\begin{equation}
      \Psi_{qq} = \overline{q}^a(\bx) q^a(\by) \Psi_0.
\label{zero_gluon}
\end{equation}
The $R$-dependent part of the energy expectation value of this state
is precisely the Coulomb energy.  If it is true, as our data suggests,
that (i) the Coulomb potential
is linearly confining; and (ii) the Coulomb string tension $\s_{coul}$
is greater than the string tension $\s$ of the usual static potential,
then we find that this simplest no-chain state is indeed confining,
and of higher energy than the lowest energy flux tube state, which we
suggest has the form of a gluon chain.

%
%
\section{Conclusions}

    Our numerical results, extracted from the correlators of timelike
Wilson loops in Coulomb gauge, are relevant to several ideas about
confinement.  First of all, we have found that the Coulomb energy
grows linearly with quark separation $R$ at large $R$, in agreement with
a result long maintained by Cucchieri and Zwanziger \cite{ZC,ZC1}.
We also find that
when center vortices are removed from lattice configurations by the
de Forcrand--D'Elia procedure, the Coulomb string tension $\s_{coul}$ drops
to zero.  This finding lends further support to the contention
that center vortices are crucial to the confinement property.

   Finally, our data
indicates that $\s_{coul}$ is several times larger than the
asymptotic string tension; a result which is entirely consistent
with the inequality $V(R) \le V_{coul}(R)$ at large distances, and
which bears on the validity of
the gluon chain model.  In order to have $\s_{coul} \approx \s$,
it would be necessary that the string tension $\s(T)$, extracted from
correlators of timelike lines of length $T$ and $T+1$, should be almost
independent of $T$, since $\s_{coul}=\s(0)$ at large $\b$, and
$\s=\s(\infty)$.  This is not what is found; instead $\s(T)$
drops off sharply with $T$ near $T=0$. If this result holds at
still larger values of $\b$, then we must conclude that $\s_{coul}$
is substantially higher than the usual asymptotic string
tension.\footnote{We note however that the authors of Ref.\ \cite{ZC},
who use inversion of the Faddeev--Popov operator to
compute the Coulomb string tension, reach a somewhat different
conclusion.}

     Assuming that in fact $\s_{coul}$ is greater than $\s$, it follows
that the simplest state ($\Psi_{qq}$) containing static sources but no
constituent gluons is overconfining, and therefore has a negligible
overlap with the true QCD flux tube state at very large quark separations.
A lower interaction energy must be obtainable by operating on the
vacuum with some arrangement of gluon operators, and the gluon chain model
is a specific proposal for that arrangement.  While $\s_{coul}>\s$ does
not necessarily imply that the gluon chain proposal is right, the result
$\s_{coul}\approx \s$ would have been a strong indication that the
proposal is wrong.

%
%
\begin{acknowledgments}
J.G. is pleased to acknowledge correspondence with Daniel
Zwanziger, which inspired this investigation.  Our research is
supported in part by the U.S.\ Department of Energy under Grant No.\
DE-FG03-92ER40711 (J.G.), and the Slovak Grant Agency for Science,
Grant No.\ 2/3106/2003 (\v{S}.O.).
\end{acknowledgments}

%
%

%
%
\end{document}